\documentclass[aps,pra,twocolumn]{revtex4}
\usepackage{graphicx}    
\usepackage{amsmath}

\begin{document}

\title{Dynamically generated quadrature and photon-number variances for Gaussian states}

\author{Moorad Alexanian}
\email[]{alexanian@uncw.edu}

\affiliation{Department of Physics and Physical Oceanography\\
University of North Carolina Wilmington\\ Wilmington, NC
28403-5606\\}

\date{\today}

\begin{abstract}
We calculate exactly the quantum mechanical, temporal Wigner quasiprobability density for a single-mode, degenerate parametric amplifier for a system in the Gaussian state, viz., a displaced-squeezed thermal state. The Wigner function allows us to calculate the fluctuations in photon number and the quadrature variance. We contrast the difference between the nonclassicality criteria, which is independent of the displacement parameter $\alpha$, based on the Glauber-Sudarshan quasiprobability distribution $P(\beta)$ and the classical/nonclassical behavior of the Mandel $Q_{M}(\tau)$ parameter, which depends strongly on $\alpha$. We find a phase transition as a function of $\alpha$ such that at the critical point $\alpha_{c}$, $Q_{M}(\tau)$, as a function of $\tau$, goes from strictly classical, for $|\alpha|< |\alpha_{c}|$, to a mixed classical/nonclassical behavior, for $|\alpha|> |\alpha_{c}|$.\\

\textbf{Keywords}: degenerate parametric amplifier; displaced-squeezed thermal states; Wigner function; Mandel parameter; nonclassicality criteria
\end{abstract}

\maketitle {}

\section{Introduction}

The generation of nonclassical radiation fields, e.g., quadrature-squeezed light, photon antibunching, sub-Poissonian statistics, etc., establishes the discrete nature of light and serves to study fundamental questions regarding the interaction of quantized radiation fields with matter \cite{GSA13}.

In a recent work \cite{MA16}, a detailed study was made of the temporal development of the second-order coherence function $g^{(2)}(\tau)$ for Gaussian states---displaced-squeezed thermal states---the dynamics of which is governed by a Hamiltonian for degenerate parametric amplification. The time development of the Gaussian state is generated by an initial thermal state and the system subsequently evolves in time where the usual assumption of statistically stationary fields is not made.

Nonclassicality were observed \cite{MA16} for various values of the parameters governing the temporal development of the coherence function $g^{(2)}(\tau)$---such as the coherent parameter $\alpha$, squeeze parameter $\xi$, and the mean photon number $\bar{n}$ of the initial thermal state. Our characterization of nonclassicality was based solely on the coherence function violating inequalities satisfied by the classical correlation functions.

More recently \cite{MAa16}, we dwelt into the notion of nonclassicality based on the characteristic function $\chi(\eta)$ and its two-dimensional Fourier transform to determine the existence or nonexistence of the quasiprobability distribution $P(\beta)$ of the Glauber-Sudarshan coherent or P representation of the density of state. It was shown \cite{MAa16} that the nonclassicality criteria for the radiation field, based on the one-time function $P(\beta)$, cannot characterize the classical, quantum mechanical or mixed nature of the dynamical system as attested by the temporal behavior of the two-time $g^{(2)}(\tau)$ function.

It is interesting that an analogous result was also obtained recently where it is argued that negative full counting statistics captures nonclassicality in the dynamics of the system in contrast to more conventional quasiprobability distributions that captures nonclassicality in the instantaneous state of the system \cite{HC16}.

In this paper we show that the nonclassicality criteria based on the one-time $P(\beta)$ function cannot even characterize the classical/nonclassical behavior of the one-time Mandel $Q_{M}(\tau)$ parameter as a function of $\tau$.

Generic compact expressions for the Wigner function for the one-mode electromagnetic field for general mixed Gaussian quantum states are well known \cite{DMM94}. The Wigner function is given in terms of five real parameters, viz. three of them are the variances and the covariance of photon quadrature components, while two others are the means of the quadratures. Of course, the importance of the present work is the exact time development of the dynamical system governed by a Hamiltonian for degenerate parametric amplification giving the explicit time dependence of all those five real parameters without assuming statistically stationary fields.

We consider in Sec. II the general Hamiltonian of the degenerate parametric amplifier and present the result for the quantum degree of second-order coherence $g^{(2)}(\tau)$ of Ref. \cite{MA16}. In Sec. III, we give an explicit expression of the exact, time-dependent Wigner quasiprobability density. In Sec. IV, use is made of the Wigner function to calculate the field-quadrature variance. Sec. V gives the results for the photon-number variance.  In Sec. VI, we present the differing criteria for nonclassicality. In Sec. VII, we compare numerically the behavior of the Mandel parameter and determine the existence of a phase transition. Sec. VIII summarizes our results. In Appendix A we give an explicit expression for the Wigner function in terms of the quadrature components $x_{\lambda}$ and $x_{\lambda + \pi/2}$ and express it in a form akin to the generic expression for the general mixed Gaussian quantum states. Finally, in Appendix B we present the relevant mathematical expressions that characterize the phase transition.

\section {Degenerate parametric amplification}

The Hamiltonian for degenerate parametric amplification, in the interaction picture, is
\begin{equation}
\hat{H} = c \hat{a}^{\dag 2} + c^* \hat{a}^2 + b\hat{a} + b^* \hat{a}^\dag.
\end{equation}
The radiation field is initially in a thermal state $\hat{\rho}_{0}$ and a after a preparation time $t$, the radiation field develops in time into a Gaussian state and so \cite{MA16}
\begin{equation}
\hat{\rho}_{G}=\exp{(-i\hat{H}t/\hbar)}\hat{\rho}_{0} \exp{(i\hat{H}t/\hbar)}
\end{equation}
\[
= \hat{D}(\alpha) \hat{S}(\xi)\hat{\rho}_{0} \hat{S}(-\xi) \hat{D}(-\alpha),
\]
with  the displacement $\hat{D}(\alpha)= \exp{(\alpha \hat{a}^{\dag} -\alpha^* \hat{a})}$  and the squeezing $\hat{S}(\xi)=  \exp\big{(}-\frac{\xi}{2} \hat{a}^{\dag 2} + \frac{\xi^*}{2} \hat{a}^{2} \big{ )}$ operators, where $\hat{a}$ ($\hat{a}^{\dag})$ is the photon annihilation (creation) operator, $\xi = r \exp{(i\theta)}$, and $\alpha= |\alpha|\exp{(i\varphi)}$. The thermal state is given by

\begin{equation}
\hat{\rho}_{0} = \exp{(-\beta \hbar \omega\hat{n})}/ \textup{Tr}[\exp{(-\beta \hbar \omega \hat{n})}],
\end{equation}
with $\hat{n}= \hat{a}^\dag \hat{a}$ and $\bar{n}= \textup{Tr}[\hat{\rho}_{0} \hat{n}]$ .

The parameters $c$ and $b$ in the degenerate parametric Hamiltonian (1) are determined \cite{MA16} by the parameters  $\alpha$ and $\xi$ of the Gaussian density of state (2) via
\begin{equation}
tc = -i\frac{\hbar}{2} r\exp(i\theta)
\end{equation}
and
\begin{equation}
tb= -i\frac{\hbar}{2}\Big{(} \alpha \exp{(-i\theta)} + \alpha^* \coth (r/2)\Big{)} r,
\end{equation}
where $t$ is the time that it takes the radiation field governed by the Hamiltonian (1) to generate the Gaussian density of state $\hat{\rho}_{G}$ from the initial thermal density of state $\hat{\rho}_{0}$.

The quantum mechanical seconde-order, degree of coherence is given by \cite{MA16}
\begin{equation}
g^{(2)}(\tau) = \frac{\langle \hat{a}^{\dag}(0) \hat{a}^{\dag}(\tau) \hat{a}(\tau) \hat{a}(0)\rangle }{\langle \hat{a}^{\dag}(0)  \hat{a}(0)\rangle \langle \hat{a}^{\dag}(\tau)\hat{a}(\tau) \rangle},
\end{equation}
where all the expectation values are traces with the Gaussian density operator, viz., a displaced-squeezed thermal state. Accordingly, the radiation field is initially in the thermal state $\hat{\rho}_{0}$. After time $t$, the radiation field evolves to the Gaussian state $\hat{\rho}_{G}$ and a photon is annihilated at time $t$, the system then develops in time and after a time $\tau$ another photon is annihilated \cite{MA16}. Therefore, two photon are annihilated in a time separation $\tau$ when the radiation field is in the Gaussian density state $\hat{\rho}_{G}$.

It is important to remark that we do not suppose statistically stationary fields. Therefore, owing to the $\tau$ dependence of the number of photons in the cavity in the denominator of Equation (6), $g^{(2)}(\tau)$ asymptotically, as $\tau\rightarrow \infty$, approaches a finite limit without supposing any sort of dissipative processes \cite{MA16}. The coherence function $g^{(2)}(\tau)$ is a function of $\Omega \tau=(r/t)\tau$, $\alpha$, $\xi$, and the average number of photons $\bar{n}$ in the initial thermal state (3), where the preparation time $t$ is the time that it takes the system to dynamically generate the Gaussian density $\hat{\rho}_{G}$ given by (2) from the initial thermal state $\hat{\rho}_{0}$ given by (3). Note that the limit $r\rightarrow 0$ is a combined limit whereby $\Omega =r/t$ also approaches zero resulting in a correlation function which has a power law decay as $\tau/t \rightarrow \infty$ rather than an exponential law decay as $\tau/t \rightarrow \infty$ as is the case in the presence of squeezing when $r>0$ \cite{MA16}.

\section{Wigner quasiprobability density}

The dynamics of the system is governed by the degenerate parametric amplification Hamiltonian (1) that generates the Gaussian state $\hat{\rho}_{G}$ from the initial thermal state $\hat{\rho}_{0}$ and subsequently determines the temporal behavior of the system \cite{MA16}. One has that

\begin{equation}
\hat{\rho}(t+\tau) =\exp\big{(}-i\hat{H}(t+\tau)\big{)} \hat{\rho}_{0}\exp \big{(}i\hat{H}(t+\tau)\big{)}
\end{equation}
\[
= \exp(-i\hat{H}\tau) \hat{\rho}_{G}\exp(i\hat{H}\tau).
\]
Accordingly, for any operator function $\mathcal{\hat{O}}(\hat{a},\hat{a}^\dag)$,
\begin{equation}
\textup{Tr}[\hat{\rho}(t+\tau) \mathcal{\hat{O}}(\hat{a},\hat{a}^\dag)] = \textup{Tr}[\hat{\rho}_{G} \mathcal{\hat{O}}\big{(}\hat{a}(\tau),\hat{a}^\dag (\tau)\big{)}]
\end{equation}
\[
\equiv \langle  \mathcal{\hat{O}}\big{(}\hat{a}(\tau),\hat{a}^\dag (\tau)\big{)} \rangle .
\]

One obtains for the characteristic function \cite{MAa16}
\[
\chi(\eta) = \textup{Tr}[\hat{\rho}(t+\tau)\exp{(\eta \hat{a}^\dag}-\eta^*\hat{a})]\exp{(|\eta|^2/2)}
\]
\[
=\textup{Tr}[\hat{\rho}(t+\tau)\exp{(\eta \hat{a}^\dag}) \exp{(-\eta^*\hat{a})}]
\]
\begin{equation}
=\exp{(|\eta|^2/2)} \exp{\big{(}\eta A^*(\tau)- \eta^* A(\tau)\big{)}}
\end{equation}
\[
\times \exp{\big{(}-(\bar{n}+1/2)|\xi(\tau)|^2\big{)}},
\]
where
\[
A(\tau) =\alpha \Bigg{(}\cosh(\Omega\tau)+\frac{1}{2}\coth(r/2) \sinh (\Omega \tau)
\]
\begin{equation}
-\frac{1}{2} (\cosh(\Omega \tau)-1)+\exp[i(\theta -2 \varphi)]\Big{[} -\frac{1}{2}\sinh(\Omega\tau)
\end{equation}
\[
-\frac{1}{2}\coth(r/2)\big{(}\cosh(\Omega\tau)-1\big{)}\Big{]} \Bigg{)}
\]
and
\begin{equation}
\xi(\tau)= \eta\cosh(\Omega \tau +r) +\eta^* \exp(i\theta) \sinh(\Omega \tau +r),
\end{equation}
with the displacement parameter $\alpha=|\alpha|e^{i\varphi}$, the squeezing parameter $\xi=re^{i\theta}$, and $t$ representing the time it takes the radiation field to dynamically evolve from the thermal state $\hat{\rho}_{0}$ to the Gaussian state $\hat{\rho}_{G}$.

Define
\begin{equation}
|\xi(\tau)|^2 = \eta^2 T^*(\tau) +\eta^{*2} T(\tau) +\eta \eta^* S(\tau),
\end{equation}
with
\begin{equation}
T(\tau)= \frac{1}{2} \exp{(i \theta)} \sinh [2(\Omega \tau + r)]
\end{equation}
and
\begin{equation}
S(\tau)= \cosh[2(\Omega \tau +r)].
\end{equation}

The Wigner function \cite{GSA13} is defined by
\begin{equation}
W(\beta) =\frac{1}{\pi^2} \int \textup{d}\eta^2\chi(\eta) e^{-|\eta|^2/2} \exp(-\beta^*\eta +\beta \eta^*).
\end{equation}
Note the presence of the factor $e^{-|\eta|^2/2}$ in the Wigner function (15), which is absent in the definition of the quasiprobability distribution $P(\beta)$ given in Equation (16) of Ref. \cite{MAa16}.

The integral (15) can be carried out for the characteristic function (9) and so
\begin{equation}
W(\beta)=\frac{2}{\pi} \frac{1}{\sqrt{4a^2b^2-c^2}} e^{-(a^2f^2+b^2d^2+cfd)/(4a^2b^2-c^2)},
\end{equation}
where
\[
a^2= (\bar{n}+1/2)\big{(}T(\tau) +T^*(\tau) +S(\tau)\big{)},
\]
\[
b^2=  -(\bar{n}+1/2)\big{(}T(\tau) +T^*(\tau) -S(\tau)\big{)},
\]
\begin{equation}
c=-2 i(\bar{n}+1/2)\big{(}T^*(\tau) -T(\tau)\big{)},
\end{equation}
\[
d= i(A(\tau) -A^*(\tau)-\beta+ \beta^*),
\]
\[
f= A(\tau)+A^*(\tau) -\beta-\beta^*.
\]
Note the absence of the term $-\frac{1}{2}$ in both $a^2$ and $b^2$ in (17) which terms appear in the expression for the quasiprobability distribution $P(\beta)$ as given by Equation (18) in Ref. \cite{MAa16}. The absence of such terms has quite an important consequence for the nonclassicality criteria based on the Wigner function.

The existence of a real-valued function $W(\beta)$ requires
\begin{equation}
(4a^2b^2-c^2) = 4(\bar{n}+ 1/2)^2 \geq 0,
\end{equation}
with the aid of (17), which is obviously satisfied.

The existence of $W(\beta)$ requires, owing to the normalization condiition $\int W(\beta) \textup{d}\beta^2=1$, that  $W(\beta) \rightarrow 0$ as $|\beta|\rightarrow \infty$. The bilinear form $(a^2f^2+b^2d^2+cfd)$ in the exponential in (16) can be diagonalized in the variables $\Re{(A(\tau)-\beta)}$  and $\Im{(A(\tau)-\beta)}$ resulting in the eigenvalues $(\bar{n}+1/2)\exp\big{(}-2(\Omega\tau +r)\big{)}$  and $(\bar{n}+1/2)\exp\big{(}2(\Omega\tau +r)\big{)}$ that must be nonnegative, which is obviously so for $0\leq \tau< \infty$. Accordingly, the Wigner function $W(\beta)>0$. The positive definiteness of $W(\beta)$ does not preclude, however, quantum behavior in the field-quadrature and photon number variances.

The displaced thermal state follows directly from (16) in the limit $r \rightarrow 0$, where $\Omega =r/t \rightarrow 0$, and so $W(\beta)=(1/(\pi (\bar{n}+1/2))\exp{(-|\beta-\alpha|^2/(\bar{n}+1/2))}$. The result for the coherent state follows with $\bar{n}=0$ and for the thermal state with $\alpha=0$.

One can express the Wigner function $W(\beta)$ in terms of quadratures of the field and so
\[
W(x_{\lambda}, x_{\lambda+\pi/2})=\frac{1}{2\pi^2}\int_{-\infty}^{\infty} \textup{d}\sigma\int_{-\infty}^{\infty} \textup{d}\kappa   e^{i(\sigma x_{\lambda}+ \kappa x_{\lambda +\pi/2})}
\]
\begin{equation}
\times \textup{Tr}[\hat{\rho}(t+\tau)e^{-i(\sigma\hat{x}_{\lambda} +\kappa \hat{x}_{\lambda+\pi/2})}]
\end{equation}
\[
= \frac{1}{\pi} \frac{1}{\sqrt{4a^2b^2-c^2}}e^{-(a^2f^2+b^2d^2+cfd)/(4a^2b^2-c^2)},
\]
where the factor $1/\pi$ in (19), as compared to the factor $2/\pi$ in (16), is a direct consequence of the change of integration variables $\eta= (\kappa -i\sigma)e^{i\lambda} /\sqrt{2}$ from (15) to (19) whereby $(\textup{d} \Re \eta)(\textup{d}\Im \eta)=\frac{1}{2}(\textup{d}\sigma)(\textup{d}\kappa)$. The quadrature variables are
\begin{equation}
x_{\lambda}=\frac{\beta e^{-i\lambda}+\beta^* e^{i\lambda}}{\sqrt{2}} \hspace{0.3in}x_{\lambda+\pi/2}=\frac{\beta e^{-i\lambda}-\beta^* e^{i\lambda}}{\sqrt{2}i},
\end{equation}
where
\begin{equation}
\beta + \beta^*= \sqrt{2}( x_{\lambda} \cos\lambda - x_{\lambda +\pi/2}\sin\lambda )
\end{equation}
and
\begin{equation}
\beta - \beta^*= \sqrt{2}i( x_{\lambda} \sin\lambda + x_{\lambda +\pi/2}\cos\lambda )
\end{equation}
which are substituted in the expressions for $d$ and $f$ in Equation (17) when evaluating (19).

Accordingly, the probability distribution for the two quadrature components $x_{\lambda}$ and $x_{\lambda+\pi/2}$ is given by $W( x_{\lambda}, x_{\lambda+\pi/2})$, which is a Gaussian function in both variables, that is, the exponential factor in (19) is a quadratic form in $x_{\lambda}$ and $x_{\lambda+\pi/2}$ . [See Equation (A9) in Appendix A.]

\section{Field-quadrature variance}

With the aid of successive derivatives of the characteristic function $\chi(\eta)$, one obtains for the quadrature $\langle\hat{x}_{\lambda}\rangle$ and the quadrature variance $\Delta x^2_{\lambda}$
\begin{equation}
\langle \hat{x}_{\lambda}\rangle = \textup{Tr}[\hat{\rho}(t+\tau) \hat{x}_{\lambda}]  =  \frac{1}{\sqrt{2}}[A(\tau) e^{-i\lambda} + A^*(\tau) e^{i\lambda}]
\end{equation}
and
\[
\Delta x^2_{\lambda} =\textup{Tr}[\hat{\rho}(t+\tau)( \hat{x}_{\lambda} -\langle \hat{x}_{\lambda}\rangle)^2 ]
\]
\begin{equation}
=(\bar{n} + 1/2) \Big{(}\exp{[2(\Omega \tau +r)]}\sin^2(\lambda-\theta/2)
\end{equation}
\[
+  \exp[-2(\Omega \tau +r)] \cos^2(\lambda-\theta/2) \Big{ )},
\]
where the quadrature operator $\hat{x}_{\lambda} = (\hat{a}e^{-i \lambda} + \hat{a}^\dag e^{i \lambda})/\sqrt{2}$. The phase-sensitive quadrature operators represent a set of observables that can be measured for radiation modes, atomic motion in a trap, and other related systems \cite{WVO99}.

The average (23) and variance (24) can also be evaluated with the aid of the probability distribution $W( x_{\lambda}, x_{\lambda+\pi/2})$.

The expectation value of $\hat{x}_{\lambda}$ is determined by the coherent amplitude  $\alpha$ as well as the squeezing parameter $\xi$ while the variance $\Delta x^2_{\lambda}$, and hence the squeezing, depends on the squeezing parameter $\xi$ only. The product of the variances of the two quadratures components $\hat{x}_{\lambda}$ and $\hat{x}_{\lambda + \pi/2}$ is bounded from below by the Heisenberg uncertainty principle since
\begin{equation}
(\Delta x^2_{\lambda} )(\Delta x^2_{\lambda+ \pi/2})= (\bar{n}+1/2)^2  \big{(}\cosh^2[2(\Omega \tau +r)]
\end{equation}
\[
- \cos^2(\theta- 2 \lambda) \sinh^2[2(\Omega \tau +r)]\big{)}\geq (\bar{n}+1/2)^2 \geq  \frac{1}{4},
\]
where the first inequality becomes an equality for $\theta=2\lambda$.

The signal-to-noise ratio \cite{RL00} is defined as
\begin{equation}
\textup{SNR} = \frac{\langle \hat{x}_{\lambda}\rangle^2}{\Delta x^2_{\lambda}}.
\end{equation}
Thus the maximum signal-to-noise ratio is
\begin{equation}
\textup{SNR}_{\textup{max}} =|\alpha|^2 \frac{\big{[} \coth(r/2)(1- e^{-\Omega \tau}) +(1 +e^{-\Omega \tau})\big{]}^2}{(2\bar{n} +1) e^{-2(\Omega \tau +r)}},
\end{equation}
for $\varphi =\lambda= \theta/2$. The result for the squeezed coherent state, $4 e^{2r}|\alpha|^2$, follows for $\tau=0$ and $\bar{n}=0$.

\section{ Photon-number variance}

Similarly, with the aid of successive derivatives of the characteristic function $\chi(\eta)$, the time development of the photon number is given by
\[
\textup{Tr}[\hat{\rho}(t+\tau)\hat{a}^\dag \hat{a}] = \langle \hat{a}^\dag(\tau) \hat{a}(\tau)\rangle = \langle\hat{n}(\tau)\rangle
\]
\begin{equation}
=(\bar{n}+1/2)\cosh[2(\Omega\tau+r)] + |A(\tau)|^2 -\frac{1}{2},
\end{equation}
while the variance is
\[
\Delta n^2(\tau) = \textup{Tr}[\hat{\rho}(t+\tau) (\hat{n} - \langle \hat{n}\rangle)^2] = (\bar{n}+1/2)^2 \cosh[4(\Omega \tau +r)]
\]
\begin{equation}
 +(\bar{n}+1/2)\Big{(} 2\cosh[2(\Omega \tau +r)]|A(\tau)|^2 -\sinh[2(\Omega \tau+r)]
\end{equation}
\[
\times [e^{i\theta} A^{*2}(\tau)+e^{-i\theta} A^2(\tau)]\Big{)} -\frac{1}{4}.
\]

Note, contrary to the quadrature variance (24), the photon-number variance (29) depends, in addition to the squeezing parameter $\xi$,  also on the coherent amplitude $\alpha$ via $A(\tau)$ given by Equation (10).

\section{Nonclassicality criteria}

A sufficient conditions for nonclassicality is for the quadrature of the field to be narrower than that for a coherent state, that is,
\begin{equation}
\Delta x^2_{\lambda} < \frac{1}{2}.
\end{equation}

Another sufficient condition is determined by the Mandel $Q_{M}(\tau)$ parameter related to the photon-number variance \cite{GSA13}
\begin{equation}
Q_{M}(\tau)= \frac{\Delta n^2(\tau) -\langle \hat{n}(\tau)\rangle}{\langle \hat{n}(\tau)\rangle},
\end{equation}
where $-1 \leq Q_{M}(\tau) <0$ implies that the field must be nonclassical with sub-Poissonian statistics. In the Glauber-Sudarshan coherent state or P-representation, nonclassicality is signaled by the real function $P(\beta)$ assuming negative values or becoming more singular than a Dirac delta function. Note, however, that if both the Mandel $Q_{M}(\tau)$ parameter and the squeezing parameter $(\Delta x_{\lambda}^2-1/2)$ are positives, then no conclusion can be drawn on the nonclassical nature of the radiation field \cite{GSA13}.

The necessary and sufficient condition \cite{MAa16} for the existence of a real-valued $P(\beta)$ is
\begin{equation}
1\leq (2\bar{n}+1)e^{-2(\Omega \tau +r)}.
\end{equation}
Accordingly, the necessary and sufficient condition for nonclassicality is then
\begin{equation}
(2\bar{n}+1)e^{-2(\Omega \tau +r)}<1,
\end{equation}
which is the same, with the aid of (24), as that given by condition (30) when $\theta= 2 \lambda$. If the nonclassicality condition (33) holds for $\tau=0$, then it holds for $\tau>0$. Therefore, the radiation field, if initially nonclassical remains so as time goes on. If the field is initially classical, that is, $(2\bar{n} +1) e^{-2r}\geq 1$, then for $\Omega \tau > \frac{1}{2} \ln [(2\bar{n} +1) e^{-2r}]$ the radiation field behaves nonclassically.

Note, however, that $Q_{M}(\tau)$ need not mirror the classical/nonclassical behavior dictated by criteria (32)-(33) based on $P(\beta)$. [See Figures. (1) and (2) below.]

\section{Numerical comparisons}

Owing to the equivalence for $\theta=2\lambda$ of the nonclassical condition $\triangle x^2_\lambda <1/2$ given by (30) and the nonclassicality criteria (33) based on the Glauber-Sudarshan $P(\beta)$ function, we need study only numerically the relation of classicality or nonclassicality between the conditions based on the $P(\beta)$ and the $Q_{M}(\tau)$ functions.

It is interesting that Equation (33) is independent of the coherent parameter $\alpha$ while the Mandel parameter $Q_{M}(\tau)$ is rather sensitive to the value of $\alpha$. This is so since the eigenvalues associated with the quadratic form, appearing in the exponential of the Wigner function $W(x_{\lambda}, x_{\lambda +\pi/2})$ in (19), do not depend on $A(\tau)$ and so are independent of the value of the displacement parameter $\alpha$. However, both the photon-number variance $\triangle n^2(\tau)$ given by (29) and the average photon number (28) do depend on the values of $\alpha$ and so does $Q_{M}(\tau)$.

Figure 1 illustrates the case $(2\bar{n}+1)e^{-2r}\geq1$ where $Q_{M}(0)>0$ for all values of $\bar{n}$ and $r$ that satisfy inequality (B2), which indicates that the system behaves classically at $\Omega \tau=0$. The behavior of $Q_{M}(\tau)$ as a function of $|\alpha|$ shows how the system goes from a strictly classical  behavior for $|\alpha| < |\alpha_{c}|=0.3494$ (blue graph) as a function of $\tau$ to a mixed classical/nonclassical behavior for $|\alpha| > |\alpha_{c}|$ (green graph) as a function of $\tau$. This behavior of $Q_{M}(\tau)$ characterizes a phase transition as a function of $|\alpha|$ when the system goes from a strictly classical behavior to one where the system exhibits, as a function of $\tau$, both classical and nonclassical behaviors. This behavior of the Mandel parameter $Q_{M}(\tau)$ as a function of $|\alpha|$ is reminiscent of the Van der Waals equation of state with the aid of the Maxwell construction whereby there exists  a critical temperature, where $1/|\alpha_{c}|$ play the role of the critical temperature $T_{c}$, above which the system is in a single phase and below which there are two coexisting phases.

\begin{figure}
\begin{center}
   \includegraphics[scale=0.3]{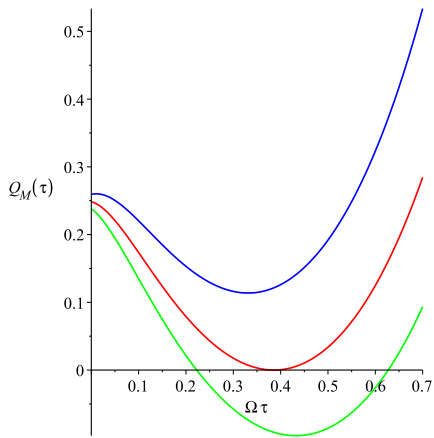}
\end{center}
\label{fig:theFig}
  \caption{Plot of $Q_{M}(\tau)$ for $\bar{n}=0.2$ and $r=0.1$ with $(2\bar{n}+1)e^{-2r}=1.1462>1$. (a) $|\alpha|=0.3$ (blue graph). (b) $|\alpha|=|\alpha_{c}|=0.3494$, where $\textup{d}Q_{M}(\tau)/\textup{d}\tau =0$ and $Q_{M}(\tau)=0$ at $\Omega \tau= 0.3857$ (red graph). (c) $|\alpha|=0.4$ (green graph).}
\end{figure}

We consider in Figure 2 the case $(2\bar{n}+1)e^{-2r}<1$, where $Q_{M}(0)$, for given $\bar{n}$ and $r$, can assume either positive or negative values as determined by inequalities (B2) and (B3), respectively, and $Q_{M}(0)=0$ if the inequalities hold as equalities. Accordingly, depending on the value of $|\alpha|$, for given $\bar{n}$ and $r$, the system behaves classically or nonclassically at $\Omega \tau =0$. There is a critical value $|\alpha_{c}|$ such that for $|\alpha|<|\alpha_{c}|$, the system behaves strictly classically (green and blue graphs); whereas for $|\alpha|>|\alpha_{c}|$ the system is in a mixed classical/nonclassical state (red and black graphs). In contrast with the results of Figure 1, for $(2\bar{n}+1)e^{-2r} \geq 1$, in the case of Figure 2, for $(2\bar{n}+1)e^{-2r} <1$, we have an analogous transition at the critical point $|\alpha_{c}|$ but with the sign of $Q_{M}(0)$ changing from positive to negative values at the critical point. Note that $Q_{M}(0)$ remains positive for the case of Figure 1 as the transition point is traversed.

\begin{figure}
\begin{center}
   \includegraphics[scale=0.3]{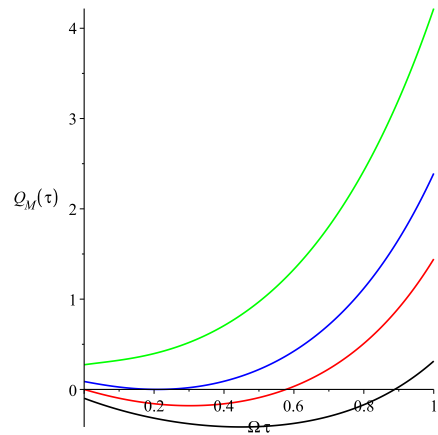}
\end{center}
\label{fig:theFig}
  . \caption{Plot of $Q_{M}(\tau)$ for $\bar{n}=0.1$ and $r=0.2$ with $(2\bar{n}+1)e^{-2r}=0.8044<1$. (a) $|\alpha|=0.3$ (green graph). (b) $|\alpha|=|\alpha_{c}|=0.4961$, where $\textup{d}Q_{M}(\tau)/\textup{d}\tau =0$ and $Q_{M}(\tau)=0$ at $\Omega \tau= 0.2097$ (blue graph). (c) $|\alpha|=0.6507$, where $Q_{M}(0)=0$ (red graph). (d) $|\alpha|=1$ (black graph).}
\end{figure}

Figure 3 shows sequence of plots for $\bar{n}=1$ and $r=1$ as the amplitude $|\alpha|$ increases past the critical value $|\alpha_{c}|=9.7140$ whereby the Mandel parameter $Q_{M}(\tau)$, as a function of $\tau$, goes from exhibiting a strictly classical behavior for $|\alpha|<|\alpha_{c}|$ to a mixed classical/nonclassical behavior for $|\alpha|>|\alpha_{c}|$.

It is important to remark that even though $Q_{M}(\tau)$ exhibits classical behavior, nonetheless, the radiation field is nonclassical since $(2\bar{n}+1)e^{-2r} <1$. There is no inconsistency since the negativity of $Q_{M}(\tau)$ is a sufficient condition for the field to be nonclassical, whereas, if $Q_{M}(\tau)>0$, no conclusion can be drawn about the nonclassicality of the radiation field \cite{GSA13}.

\begin{figure}
\begin{center}
   \includegraphics[scale=0.3]{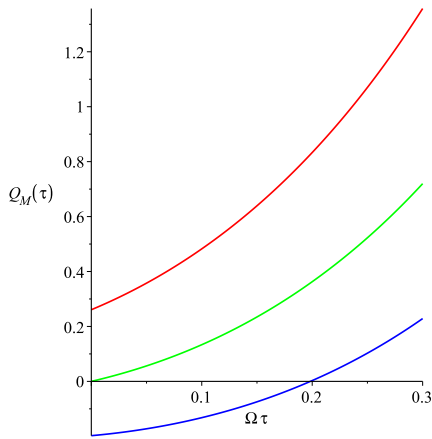}
\end{center}
\label{fig:theFig}
  . \caption{Plot of $Q_{M}(\tau)$ for $\bar{n}=1$ and $r=1$ with $(2\bar{n}+1)e^{-2r}=0.4060<1$. (a) $|\alpha|=12$ (red graph). (b) $|\alpha|=|\alpha_{c}|=9.7140$, where $Q_{M}(0)=0$  (green graph). (c) $|\alpha|=8$ (blue graph). }
\end{figure}

\section{Summary and discussions}

We calculate the Wigner quasiprobability density (19) and the corresponding field-quadrature (24) and photon-number variances (29) for Gaussian states, \emph{viz}., displaced-squeezed thermal states, where the dynamics is governed solely by the general, degenerate parametric amplification Hamiltonian (1). Our result (19) for the Wigner function is exact and is based on dynamically generating the Gaussian state first from an initial thermal state and subsequently determining the time evolution of the system without assuming statically stationary fields.

We numerically analyze the conditions for nonclassicality as given by the Mandel parameter  $-1\leq G_{M}(\tau)<0$, squeezing parameter $(\Delta x_{\lambda}^2-1/2)<0$, and the nonclassicality criteria (33) based on the Glauber-Sudarshan quasiprobability distribution $P(\beta)$. We show that the latter condition by itself, albeit determining the nonclassicality of the radiation field, does not determine the classical/nonclassical behavior exhibited by the Mandel parameter $Q_{M}(\tau)$.

The nonclassicality criteria (33) for the radiation field based on $P(\beta)$ is independent of the value of the displacement parameter $\alpha$. However, the Mandel parameter $Q_{M}(\tau)$ has a sensitive dependence on the value of $\alpha$. We find a phase transition from classical to a mixed classical/nonclassical behavior for $Q_{M}(\tau)$, as a function of $\tau$, at $|\alpha_{c}|$ so that for $|\alpha|< |\alpha_{c}|$, the behavior of  $Q_{M}(\tau)$ is strictly classical, whereas for $|\alpha|> |\alpha_{c}|$, $Q_{M}(\tau)$ exhibits both classical and nonclassical behaviors even though the radiation field is strictly nonclassical.

\appendix
\section{Wigner function}

In this appendix we express the quadratic form appearing in the exponential function in (19) explicitly in terms of the quadrature components $x_{\lambda}$ and $x_{\lambda +\pi/2}$. Now
\[
E(x_{\lambda}, x_{\lambda +\pi/2}) \equiv a^2f^2+b^2d^2+cfd
\]
\begin{equation}
=\epsilon_{(x_{\lambda},x_{\lambda})} (x_{\lambda}-\langle\hat{x}_{\lambda}\rangle)^2+\epsilon_{(x_{\lambda+\pi/2},x_{\lambda+\pi/2})}(x_{\lambda+\pi/2}-\langle\hat{x}_{\lambda+\pi/2}\rangle)^2
\end{equation}

\[
+\epsilon_{(x_{\lambda},x_{\lambda+\pi/2})} (x_{\lambda}-\langle\hat{x}_{\lambda}\rangle )(x_{\lambda+\pi/2}-\langle\hat{x}_{\lambda+\pi/2} \rangle),
\]
where
\begin{equation}
\langle \hat{x}_{\lambda}\rangle= \frac{1}{\sqrt{2}}[A(\tau)e^{-i\lambda} +A^*(\tau)e^{i\lambda}],
\end{equation}
\begin{equation}
\langle \hat{x}_{\lambda+\pi/2}\rangle= \frac{1}{i\sqrt{2}}[A(\tau)e^{-i\lambda} -A^*(\tau)e^{i\lambda}],
\end{equation}
\[
\epsilon_{(x_{\lambda},x_{\lambda})}=2(\bar{n}+1/2)\big{(}\cos(2\lambda-\theta)\sinh[2(\Omega \tau+r)]
\]
\begin{equation}
+\cosh[(2(\Omega\tau+r)]\big{)},
\end{equation}
\[
\epsilon_{(x_{\lambda+\pi/2},x_{\lambda+\pi/2})}= -2(\bar{n}+1/2)\Big{(}\cos(\theta-2\lambda)\sinh[2(\Omega\tau+r)]
\]
\begin{equation}
-\cosh[2(\Omega \tau+r)]\Big{)},
\end{equation}
and
\begin{equation}
\epsilon_{(x_{\lambda},x_{\lambda+\pi/2})}= 4(\bar{n}+1/2)\sin(\theta-2\lambda)\sinh[2(\Omega\tau+r)].
\end{equation}

One has that the Wigner quasiprobability density is given by
\begin{equation}
W(x_{\lambda}, x_{\lambda +\pi/2})= \frac{1}{\pi(\bar{n}+1/2)}  e^{-E(x_{\lambda}, x_{\lambda +\pi/2})/(2\bar{n} +1)^2},
\end{equation}
where the five real, time-dependent parameters (A2)-(A6) can be directly compared to the five real parameters  appearing in the generic Gaussian Wigner function \cite{DMM94}.

The probability distribution $P(x_{\lambda})$ for the quadrature component $\hat{x}_{\lambda}$ is
\[
P(x_{\lambda}) = \int_{-\infty}^{\infty} \textup{d}x_{\lambda+\pi/2} W(x_{\lambda}, x_{\lambda +\pi/2})
\]
\begin{equation}
=\frac{1}{\sqrt{2\pi\triangle x^2_{\lambda}}} \exp{\Big{(}-\frac{(x_{\lambda}- \langle \hat{x}_{\lambda}\rangle)^2}{2\triangle x^2_{\lambda}}}\Big{)},
\end{equation}
where $\langle \hat{x}_{\lambda}\rangle$ and $\triangle x^2_{\lambda}$ are given by (A2) and (24), respectively.

Result (A1) for $E(x_{\lambda}, x_{\lambda +\pi/2})$ simplifies considerably for the case of maximum signal-to-noise ratio when $\theta=2\lambda$ and the cross term $\epsilon_{(x_{\lambda},x_{\lambda+\pi/2})}$ in (A6) vanishes, in which case
\[
W(x_{\lambda}, x_{\lambda +\pi/2})=\frac{1}{\pi(2\bar{n}+1)}\exp{\Big{(}-\frac{(x_{\lambda}- \langle \hat{x}_{\lambda}\rangle)^2}{(2\bar{n} +1)e^{-2(\Omega \tau +r)}}} \Big{)}
\]
\begin{equation}
\times \exp{\Big{(}-\frac{(x_{\lambda+\pi/2}- \langle \hat{x}_{\lambda+\pi/2}\rangle)^2}{(2\bar{n} +1)e^{2(\Omega \tau +r)}}} \Big{)}.
\end{equation}

\section{phase transition}

The Mandel parameter $Q_{M}(0)$ follows from (31)
\[
Q_{M}(0)=\Big{(}(\bar{n} +1/2)^2 \cosh(4r)+\big{(}(2\bar{n}+1)e^{-2r}-1\big{)}|\alpha|^2
\]
\begin{equation}
-(\bar{n}+1/2)\cosh(2r) +1/4\Big{)}
\end{equation}
\[
\times \frac{1}{(\bar{n}+1/2)\cosh(2r)+|\alpha|^2 -1/2}
\]
with $\theta=2\lambda$. The value of $Q_{M}(0)$ can assume either positive or negative values.

If $Q_{M}(0)>0$, then
\begin{equation}
[1-(2\bar{n}+1)e^{-2r}]|\alpha|^2
\end{equation}
\[
< (\bar{n}+1/2)^2 \cosh(4r)-(\bar{n}+1/2)\cosh(2r) +1/4.
\]
Note that if $1-(2\bar{n}+1)e^{-2r}\leq 0$, then $Q_{M}(0)>0$ for all values of $|\alpha|$, $\bar{n}$ and $r$ provided $\bar{n}\geq(e^{2r} - 1)/2$. On the other hand, if $1-(2\bar{n}+1)e^{-2r}>0$, then $|\alpha|$ must satisfy inequality (B2).

However, if $Q_{M}(0)<0$, then
\begin{equation}
[1-(2\bar{n}+1)e^{-2r}]|\alpha|^2
\end{equation}
\[
> (\bar{n}+1/2)^2 \cosh(4r)-(\bar{n}+1/2)\cosh(2r) +1/4,
\]
which requires $1>(2\bar{n}+1)e^{-2r}$.

There are three possible behaviors of $Q_{M}(\tau)$ as a function of $\Omega \tau$ when $Q_{M}(0)>0$, (i) $Q_{M}(\tau)=0$ for two different values of $\Omega \tau$; (ii) $Q_{M}(\tau)$ vanishes for a single value of $\Omega \tau$, at which both $Q_{M}(\tau)$ and $\partial Q_{M}(\tau)/\partial \tau$ vanish; (iii) $Q_{M}(\tau) >0$ for $\Omega \tau\geq 0$.

There is only one possible behavior for the system when $Q_{M}(0)<0$, $Q_{M}(\tau)$ as a function of $\tau$ crosses the axis once when it assumes a value of zero. Therefore, the Mandel parameter is negative and so nonclassical and remains so but as time goes the Mandel parameter becomes positive and so classical. Note that the radiation field is nonclassical since inequality (33) is satisfied but this does not preclude that $Q_{M}(\tau)$ can behave classically as is shown in both Figures 2 and 3.

\begin{newpage}
\bibliography{}

\end{newpage}
\end{document}